\documentclass[aps,pre,onecolumn]{revtex4}
\usepackage{graphicx, bm, bbm,amsmath, amssymb, epsfig,color}

\pdfoutput=1

\newcommand{\bq}{\begin{align}}
\newcommand{\eq}{\end{align}}

\begin{document}
\title{Transition to branching flows in optimal planar convection}

\author{Silas Alben$^*$}
\affiliation{Department of Mathematics, University of Michigan,
Ann Arbor, MI 48109, USA}
\email{alben@umich.edu}

\date{\today}
\begin{abstract}
We study steady flows that are optimal for heat transfer in a two-dimensional periodic domain. The flows maximize heat transfer under the constraints of incompressibility and a given energy budget (i.e. mean viscous power dissipation). Using an unconstrained optimization approach, we compute optima starting from 30--50 random initializations across several decades of Pe, the energy budget parameter. At Pe between 10$^{4.5}$ and 10$^{4.75}$, convective rolls with U-shaped branching near the walls emerge. They exceed the heat transfer of the simple convective roll optimum
at Pe between 10$^{5}$ and 10$^{5.25}$. At larger Pe, multiple layers of branching occur in the optima, and become increasingly elongated and asymmetrical. Compared to the simple convective roll, the branching flows have lower maximum speeds and thinner boundary layers, but nearly the same maximum power density.
\end{abstract}


\maketitle

\section{Introduction}
The transfer of heat between solid surfaces and adjacent fluid flows is essential to many natural and technological phenomena \cite{bird2007transport,bergman2011fundamentals,lienhard2013heat,dauxois2021confronting}. Fundamental results in the field of heat transfer concern how the rate of heat transfer (quantified by the Nusselt number Nu) depends on the form of the fluid flow, with the flow magnitude quantified by the Rayleigh number (for buoyancy-driven, or free convection) or the P\'{e}clet number, Pe (for forced convection) \cite{bergman2011fundamentals,marcotte2018optimal}. Scaling laws have been derived for fluid flows that are simple in form (e.g. Poiseuille flow through a channel \cite{shah2014laminar}) or flows driven by simple forcings (e.g. pressure-driven flows past obstacles \cite{bergman2011fundamentals,glezer2016enhanced,kuwata2022dissimilar,pirozzoli2023direct} and buoyancy-driven flows between surfaces at fixed temperatures \cite{zhou2010prandtl,sondak2015optimal,toppaladoddi2017roughness,zhu2018transition,wen2020steady,wang2022off,solano2022natural}).

Recently Hassanzadeh, Chini, and Doering studied convective flows in a 2D fluid layer between horizontal walls at fixed temperatures, hot at the bottom and cold at the top, as in Rayleigh-B\'{e}nard convection \cite{hassanzadeh2014wall}. Instead of buoyancy-driven flows, they considered forced
incompressible flows and
performed a continuation search for flows that maximize the rate of heat transfer from the solid surfaces, for a given mean kinetic energy or rate of viscous dissipation of the flow (given as a type of P\'{e}clet number). They did not explicitly identify such flows as forced flows, but any incompressible flow is a solution to the Navier-Stokes equation with a given distribution of forcing per unit volume, spread over the flow region \cite{alben2017improved,alben2017optimal}. They studied steady 2D flows, periodic in the horizontal direction (along the walls), with free-slip boundary conditions. They found optimal flows in the form of rectangular convection rolls of decreasing horizontal width and sharpening boundary layers near the walls as the average input power (Pe) increases. 
Later, Souza and Doering performed a continuation search in the fixed-viscous-dissipation case with no-slip instead of free-slip boundary conditions at high Pe with higher numerical resolution and obtained Nu $\sim$ Pe$^{6/11}$, instead of Pe$^{10/17}$ in the free-slip case \cite{souza2020wall}. Tobasco and Doering analytically constructed self-similar branching flows that obtain Nu $\sim$ Pe$^{2/3}/(\log{\mbox{Pe}})^{4/3}$ \cite{tobasco2017optimal,doering2019optimal}, close to the 
$O($Pe$^{2/3})$ upper bound obtained using the background method \cite{souza2016optimal}, which holds independently of the flow boundary conditions. Meanwhile, Motoki et al. studied the same problem in three dimensions, periodic in the two directions perpendicular to the walls \cite{motoki2018maximal}. They used Newton's method to solve the system of nonlinear equations obtained by taking variations of the Lagrangian function for the heat transfer with Lagrange multipliers for the constraints of flow incompressibility, the advection-diffusion equation, and the fixed rate of viscous dissipation. At Pe = 79.2, the optimal flows bifurcate from the 2D rolls of \cite{souza2020wall} to 3D flows that eventually adopt branching structures near the no-slip walls. The branching occurs over a wider range of spatial scales as Pe increases from 500 to 10000. Along with the 3D branching structures, the scaling Nu $\sim$ Pe$^{2/3}$ was found. The same method was used earlier by the same authors to study similar transitions with Couette-flow (moving-wall) boundary conditions \cite{motoki2018optimal}. 

The goal of this paper is to search broadly for optimal 2D no-slip flows. \cite{souza2020wall} found a sequence of rectangular convective rolls as Pe was increased, while 
\cite{tobasco2017optimal,doering2019optimal} constructed a self-similar branching flow that is close to optimal in the large-Pe limit. Are there optimal flows that interpolate between these two, and are there other types of optimal flows? Is there a succession of branching flows similar to those found in 3D by \cite{motoki2018maximal}? We address these questions with a different computational method than the previous works, simpler in some ways because it uses an unconstrained optimization formulation.

\section{Model}
A layer of fluid is contained between walls at $y = 0$ and 1, with temperatures 1 and 0 respectively. We have chosen to nondimensionalize all quantities with units of length (including $y$) by the channel height $H$ and quantities with units of temperature by the temperature difference between the walls $\Delta T$.
We search for a steady 2D flow field $(u,v) = \mathbf{u} = (\partial_y\psi,-\partial_x\psi)$, automatically incompressible for a given stream function $\psi(x,y)$. The flow is assumed periodic in $x$ with period $L_x$, to be determined as part of the optimization of $\psi$. The flow velocity is zero at the walls: $\psi = \partial_y\psi = 0$ at $y = 0$ and $\psi = \psi_{top}$ and $\partial_y\psi = 0$ at $y = 1$. The constant $\psi_{top}$ is the net horizontal fluid flux through the channel. In test runs our computational solutions always converge to the case $\psi_{top} = 0$, so we assume this value in general because it reduces the number of optimization parameters. 

The temperature field is determined by the steady advection-diffusion equation 
\begin{equation}
    \mathbf{u}\cdot\nabla T - \nabla^2T = 0. \label{AdvDiff}
\end{equation}
We have set the prefactor of $\nabla^2T$ (the thermal diffusivity) to unity, corresponding to nondimensionalizing velocities by $\kappa/H$, with $\kappa$ the dimensional thermal diffusivity.
The setup is similar to Rayleigh-B\'{e}nard convection, but instead of solving the Boussinesq equations for the flow field, we
determine the flow field that maximizes the Nusselt number, the mean heat transfer from the hot lower boundary
\begin{equation}
\mbox{Nu} \equiv \frac{1}{L_x}\int_0^{L_x} -\partial_y T|_{y = 0}\; dx. \label{NuDef}
\end{equation}
We maximize Nu over flows with a given spatially-averaged rate of viscous dissipation Pe$^2$:
\begin{equation}
    \mbox{Pe}^2 \equiv \frac{1}{2}\frac{1}{L_x}\int_0^{Lx}\int_0^1 \left(\nabla \mathbf{u} + \nabla \mathbf{u}^T\right)^2 dx \,dy = \frac{1}{L_x}\int_0^{Lx} \int_0^1  \left(\nabla^2 \psi\right)dx \,dy. \label{Power}
\end{equation}
Pe$^2$ is the average rate of viscous dissipation nondimensionalized by $\mu \kappa^2/H^4$, with $\mu$ the viscosity.
The second equality in (\ref{Power}) holds for incompressible flows given the no-slip and periodic boundary conditions we have here \cite{lamb1932hydrodynamics,Mil68}.

One way to solve this constrained optimization problem is to find stationary points of the Lagrangian \cite{hassanzadeh2014wall,motoki2018maximal,souza2020wall}, which can be written as \cite{alben2017optimal,alben2017improved}:
\begin{equation}
    \mathcal{L} = \frac{1}{L_x}\int_0^{L_x} -\partial_y T|_{y = 0}\; dx + \frac{1}{L_x} \int_0^{L_x} \int_0^1 m(x,y)\left(
    -\nabla^\perp \psi \cdot\nabla T - \nabla^2T
    \right) dx \,dy + \lambda \left( \mbox{Pe}^2 - 
    \frac{1}{L_x}\int_0^{Lx} \int_0^1  \left(\nabla^2 \psi\right)^2 dx \,dy
    \right).
\end{equation}
By setting the variations of $\mathcal{L}$ with
respect to $T$, $\psi$, and the Lagrange multipliers $m$ and $\lambda$ to 0 and integrating by parts, we obtain a system of nonlinear equations and boundary conditions for these four variables that can be solved using Newton's method \cite{alben2017improved}:
\begin{gather}
    \frac{\delta\mathcal{L}}{\delta m} = -\nabla^\perp \psi \cdot\nabla T - \nabla^2T = 0 \label{Varm} \\ 
    \frac{\delta\mathcal{L}}{\delta T} = \nabla^\perp \psi \cdot\nabla m - \nabla^2m = 0 \label{VarT} \\ 
    \frac{\delta\mathcal{L}}{\delta \psi} = \nabla^\perp m \cdot\nabla T + 2\lambda\nabla^4 \psi = 0 \label{VarPsi}\\ 
    \frac{\delta\mathcal{L}}{\delta \lambda} =
    \mbox{Pe}^2 - 
    \frac{1}{L_x}\int_0^{Lx} \int_0^1  \left(\nabla^2 \psi\right)^2 dx\, dy= 0. \label{VarLam}
\end{gather}
The boundary conditions for $T$ and $\psi$ have been stated, and those for $m$ are the same as those for $T$. In the limit of small Pe, the solutions can be expanded in powers of Pe:
\begin{align}
T &= \mathcal{T}_0 + \mbox{Pe}\, \mathcal{T}_1 + \ldots \\
m &= m_0 + \mbox{Pe}\, m_1 + \ldots \\
\psi &= \mbox{Pe}\, \psi_1 + \ldots
\end{align}
with $m_0 = \mathcal{T}_0 = 1-y$, the purely conductive solution. Inserting the expansions into the equations (\ref{Varm})--(\ref{VarLam}), we find $m_1 = -\mathcal{T}_1$ and the system is reduced to a generalized eigenvalue problem
\begin{align}
    \partial_x\psi_1 - \nabla^2\mathcal{T}_1 &= 0 \\
    -\partial_x \mathcal{T}_1 -\lambda\nabla^4\psi_1 & = 0. 
\end{align}
The eigenmodes can be written as ($T_1$, $\psi_1$) = ($\hat{\mathcal{T}}_{1k}(y)\sin(kx+\phi), \hat{\psi}_{1k}(y)\cos(kx+\phi)$) for any real $k$ and an arbitrary phase angle $\phi$. In appendix \ref{OtherNu} we derive an alternative expression for Nu \cite{hassanzadeh2014wall}: 
\begin{equation}
\mbox{Nu}= 1 + \frac{1}{L_x}\int_0^1\int_0^{L_x}  vT\, dx \,dy
\end{equation}
which can be computed using $\mathcal{T}_1$ and $\psi_1$, whereas (\ref{NuDef}) would require us to also compute $\mathcal{T}_2$, the $O(\mbox{Pe}^2)$ contribution to $T$, because Nu = 1 + $O$(Pe$^2$), and the $O$(Pe$^2$) term determines the optimal flows. We have 
\begin{equation}
\mbox{Nu}= 1 + \frac{1}{2}\int_0^1 k \hat{\psi}_{1k}(y)\hat{\mathcal{T}}_{1k}(y) dy.
\end{equation}

\begin{figure}
    \centering
    \includegraphics[width=5in]{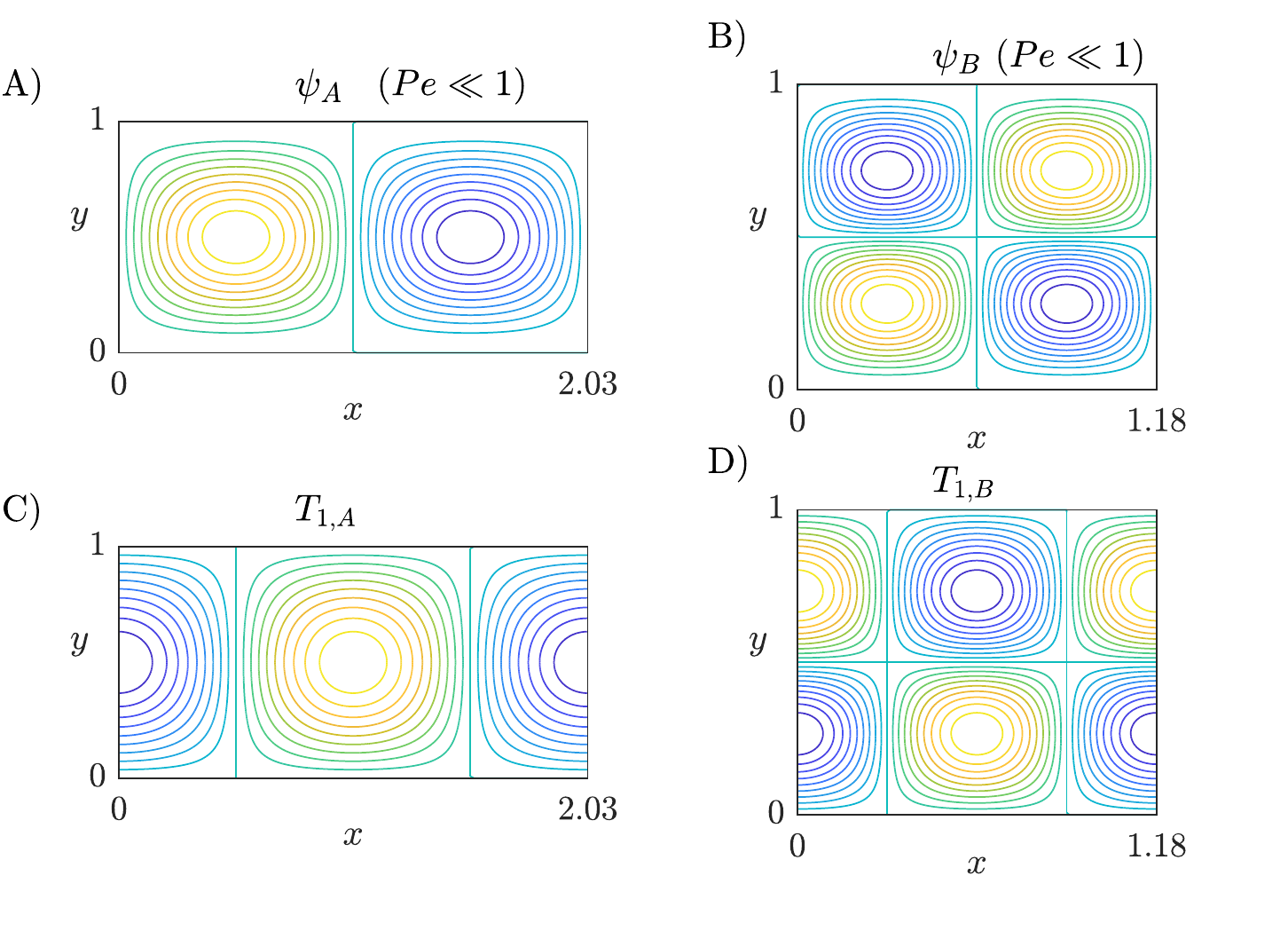}
    \caption{\footnotesize Streamlines for the best and second best optimal flows in the limit Pe $\to 0$ (A, B) and contour lines for the $O$(Pe) components of the corresponding temperature fields (C, D).}
    \label{fig:SmallPeOptima}
             \vspace{-.20in}
\end{figure}

Two examples of eigenmodes are shown in figure \ref{fig:SmallPeOptima}A,C and B,D. The flows consist of discrete sequences of convective rolls between the two walls, similar to those in \cite{hassanzadeh2014wall}, but nonsinusoidal in $y$ now due to the no-slip instead of free-slip boundary condition. The cases with one and two rolls between the walls are shown in panels A and B, and the horizontal period $L_x = 2\pi/k$ is chosen in each case to maximize Nu. Thus $\psi_A$ and $\psi_B$ are the best and second best locally optimal flows at Pe $\ll 1$, and the convective rolls are slightly larger horizontally than vertically. In panels C and D, the $O$(Pe) components of the temperature fields are shifted by 1/4 period from $\psi$ (in panels A and B), so there is a local maximum/minimum of temperature where the flow is directed vertically upward/downward between the plates.  

\cite{motoki2018maximal} and \cite{souza2020wall} solved alternative versions of the variational equations
(\ref{Varm})--(\ref{VarLam}), using $\mathbf{u}$ and the pressure $p$ instead of $\psi$, and found optimal flows at large Pe.
The system (\ref{Varm})--(\ref{VarLam}) involves fourth-order derivatives of $\psi$, and is ill-conditioned when discretized with finite differences on a fine mesh. To compute solutions at larger Pe, here we adopt an alternative approach that leads to an unconstrained optimization problem.

\section{Computational Method} 

We now describe our computational method to find optimal flows at moderate and large Pe. 
We include $L_x$ among the optimization parameters by changing variables from $x$ to $p = x/L_x$ so that the equations are solved on the fixed domain $p \in [0,1]$ and (\ref{AdvDiff})--(\ref{Power}) become
\begin{align}
    \frac{1}{L_x}\left(\partial_y \psi \partial_p T - \partial_p \psi \partial_y T\right) - \left(\frac{1}{L_x^2}\partial_p^2 + \partial_y^2\right)T= 0 \; ; \; L_x \equiv s^2\label{AdvDiffp} \\
    \mbox{Nu}= \int_0^{1} -\partial_y T|_{y = 0}\; dp. \label{NuDefp} \\
    \mbox{Pe}^2 = \int_0^{1} \int_0^1  \left(\frac{1}{L_x^2}\partial_p^2 \psi + \partial_y^2\psi\right)^2 dp\, dy. 
    \label{Powerp}
\end{align}
We optimize $L_x \equiv s^2$ by using $s$ as the optimization parameter, so $L_x$ is nonnegative while $s$ is unconstrained.
Next, we express $\psi$ as a linear combination of functions 
with period $p$ that satisfy the no-slip boundary conditions in $y$. The functions are products of Fourier modes in $p$ and
linear combinations of Chebyshev polynomials in $y$:
\begin{align}
     \psi(p,y) &= \sum_{j = 1}^{M}\sum_{k = 1}^{N-3} A_{jk}Y_k(y)\cos(2\pi p) + B_{jk}Y_k(y)\sin(2\pi p) \label{Modes} \\
     &Y_k(y) \in \langle T_0(2y-1), \ldots, T_{k+4}(2y-1) \rangle. 
\end{align}
$Y_k$ is a linear combination of Chebyshev polynomials of the first kind up to degree $k+4$ that obeys the no-slip conditions at $y = 0$ and 1. Its computation is described in appendix \ref{Numerical}. An arbitrary horizontal shift can be added to $\psi$ without changing the solution. To remove this degree of freedom we set $B_{11} = 0$.

We define a grid that is uniform in $p$ with $m = 256$ points, $\{0, 1/m, ..., 1-1/m\}$. We concentrate points near the boundaries in $y$, anticipating sharp boundary layers in the optimal flows. This is done by starting with a uniform grid for $\eta \in [0, 1]$ with spacing $1/n$, and mapping to the $y$-grid by
\begin{equation}
y = \eta -\frac{y_f}{2\pi}\sin(2\pi\eta).   \label{yeta}
\end{equation}
with $y_f$ a scalar. The $y$-spacing is maximum, $\approx (1+y_f)/n$, near $y = 1/2$, and minimum, $\approx (1-y_f)/n$, near $y = 0$ and 1. We take $y_f = 0.997$ and
$n = 512$ or 1024, giving a grid spacing $\Delta y \approx 3 \times 10^{-6}$ or $6 \times 10^{-6}$ at the boundaries.

The discretized $\psi$ is written $\mathbf{\Psi}$, a vector of values at the $m(n-1)$ interior mesh points for $(p,y) \in [0, 1) \times (0,1)$. To form $\mathbf{\Psi}$, we first normalize each mode in (\ref{Modes}) to have power Pe$^2$ (for better numerical conditioning). The differential operators and the integrals in (\ref{Powerp}) are discretized to compute the power with second-order accuracy. We arrange the modes as columns of a $m(n-1) \times 2M(N-3)-1$ matrix $\mathbf{V}$. Here we take $M = 5m/32$ and $N = 5n/32$, so we limit the modes to those whose oscillations can be resolved by the mesh. We set
$\tilde{\mathbf{\Psi}} = \mathbf{V}\mathbf{c}$, a linear combination of the discretized modes with coefficients $\mathbf{c}$.  To form a $\mathbf{\Psi}$ with power Pe$^2$, we discretize the integral in (\ref{Powerp}) as a quadratic form $\mathbf{a}^T\mathbf{M}\mathbf{a}$, where the vector $\mathbf{a}$ stands for a discretized $\psi$ in (\ref{Powerp}), and the matrix $\mathbf{M}$ gives the effect of the discretized derivatives and integrals in (\ref{Powerp}). Then
\begin{equation}
\mathbf{\Psi} = \frac{\mbox{Pe}\, \mathbf{V}\mathbf{c}}{\sqrt{(\mathbf{V}\mathbf{c})^T\mathbf{M}\mathbf{V}\mathbf{c}}} \label{Psi}
\end{equation}
has power Pe$^2$, as can be seen by evaluating $\mathbf{\Psi}^T\mathbf{M}\mathbf{\Psi}$. Such a $\mathbf{\Psi}$ automatically gives an incompressible flow by the stream function definition, and automatically satisfies the power constraint, so we are left with an unconstrained optimization problem: maximize Nu ((\ref{NuDefp}), discretized at second-order) for a given $\mathbf{c} \in \mathbb{R}^{2M(N-3)-1}$ and $s \in \mathbb{R}$.

We solve the optimization problem using the BFGS method \cite{martins2021engineering}, a quasi-Newton method that requires evaluations of the objective function Nu and its gradient with respect to the design parameters,  $\{d\mbox{Nu}/d\mathbf{c}, d\mbox{Nu}/ds\}$. Nu is computed by forming $\mathbf{\Psi}$ from $\mathbf{c}$, then computing $\mathbf{u}$ and solving (\ref{AdvDiffp}) for the discretized temperature field $\mathbf{T}$ using second-order finite differences. 

The gradient can be computed efficiently using the adjoint method \cite{martins2021engineering}. 
Using the chain rule,
\begin{equation}
    \frac{d\mbox{Nu}}{d\mathbf{c}} = \frac{d\mbox{Nu}}{d\mathbf{T}}\frac{d\mathbf{T}}{d\mathbf{\Psi}}\frac{d\mathbf{\Psi}}{d\mathbf{c}} \quad ; \quad \frac{d\mbox{Nu}}{dL_x} = \frac{d\mbox{Nu}}{d\mathbf{T}}\frac{d\mathbf{T}}{dL_x}+ \frac{d\mbox{Nu}}{d\mathbf{T}}\frac{d\mathbf{T}}{d\mathbf{\Psi}}\frac{d\mathbf{\Psi}}{dL_x}. \label{gradient}
\end{equation}
The right hand side of the second equation of (\ref{gradient}) is a sum of two terms because $\mathbf{T}$ depends on $L_x$ directly, through the coefficients of equation (\ref{AdvDiffp}), and through $\mathbf{\Psi}$, which depends on $L_x$ through $\mathbf{M}$ in the denominator of 
(\ref{Psi}). In the first equation of (\ref{gradient}), $\mathbf{T}$ depends on $\mathbf{c}$ only through $\mathbf{\Psi}$.

The entries of
$d\mathbf{T}/d\mathbf{\Psi}$ and $d\mathbf{T}/dL_x$ can be calculated using the discretized advection-diffusion equation (\ref{AdvDiffp}), written as
\begin{equation}
\mathbf{r}(\mathbf{T},\mathbf{\Psi}, L_x) = 0 \label{Residual}
\end{equation}
with $\mathbf{r}$ (the ``residual") a vector that takes values at each interior grid point. We can think of 
(\ref{Residual}) as an implicit equation for the temperature field $\mathbf{T}$ as a function of $\mathbf{\Psi}$ and $L_x$. If we apply small perturbations $(\Delta\mathbf{\Psi},\Delta L_x)$ to $\mathbf{\Psi}$ and $L_x$, we obtain a small perturbation $\Delta\mathbf{T}$ to the temperature field such that the residual remains zero:
\begin{equation}
0 = \Delta \mathbf{r} \approx \frac{d\mathbf{r}}{d\mathbf{T}}\Delta \mathbf{T} +\frac{d\mathbf{r}}{d\mathbf{\Psi}}\Delta\mathbf{\Psi} + \frac{d\mathbf{r}}{dL_x}\Delta L_x. \label{dr}
\end{equation}
In the limits $\|\Delta\mathbf{\Psi}\| \to 0$ and $\|\Delta L_x\| \to 0$, the matrices that map $\Delta\mathbf{\Psi}$ and $\Delta L_x$ to $\Delta\mathbf{T}$ are
\begin{equation}
\frac{d\mathbf{T}}{d\mathbf{\Psi}} = -\frac{d\mathbf{r}}{d\mathbf{T}}^{-1}\frac{d\mathbf{r}}{d\mathbf{\Psi}} \quad ; \quad
\frac{d\mathbf{T}}{dL_x} = -\frac{d\mathbf{r}}{d\mathbf{T}}^{-1}\frac{d\mathbf{r}}{dL_x}
 \label{dTdPsiLx}
\end{equation}
so (\ref{gradient}) becomes
\begin{equation}
    \frac{d\mbox{Nu}}{d\mathbf{c}} = -\frac{d\mbox{Nu}}{d\mathbf{T}}\frac{d\mathbf{r}}{d\mathbf{T}}^{-1}\frac{d\mathbf{r}}{d\mathbf{\Psi}}\frac{d\mathbf{\Psi}}{d\mathbf{c}} = 
    -\mathbf{q}^T\frac{d\mathbf{r}}{d\mathbf{\Psi}}\frac{d\mathbf{\Psi}}{d\mathbf{c}}\quad ; \quad
    \frac{d\mbox{Nu}}{dL_x} = -\mathbf{q}^T\frac{d\mathbf{r}}{dL_x} - \mathbf{q}^T\frac{d\mathbf{r}}{d\mathbf{\Psi}}\frac{d\mathbf{\Psi}}{dL_x}.
    \label{gradient1}
\end{equation}
where $\mathbf{q}$ is the ``adjoint variable," and is computed by solving the ``adjoint equation"
\begin{equation}
    \frac{d\mathbf{r}}{d\mathbf{T}}^T \mathbf{q} = \frac{d\mbox{Nu}}{d\mathbf{T}}^T. \label{q}
\end{equation}
The cost of solving the linear system (\ref{q}) for $\mathbf{q}$, and therefore the cost of computing the gradient terms $\{d\mbox{Nu}/d\mathbf{c}, d\mbox{Nu}/dL_x\}$, is essentially the same as that of computing Nu itself, by solving (\ref{AdvDiffp})-(\ref{NuDefp}). 
As noted above, we optimize over $s$ instead of $L_x$, to ensure $L_x \geq 0$, and use
$d\mbox{Nu}/ds$ = $2s\, d\mbox{Nu}/dL_x$ instead of $d\mbox{Nu}/dL_x$ for the gradient.

We initialize with 30-50 random choices of $\mathbf{c}$ and $s$ and run the optimization until convergence, when 
$\|(d\mbox{Nu}/d\mathbf{c}, d\mbox{Nu}/ds)\|$ is less than a threshold that ranges from $10^{-8}$ to $10^{-2}$. Larger thresholds are allowed at larger Pe, where the gradient is larger in the initial random state, and where convergence to very small gradients is more difficult.

\section{Computational results}

\begin{figure}
    \centering
    \includegraphics[width=7in]{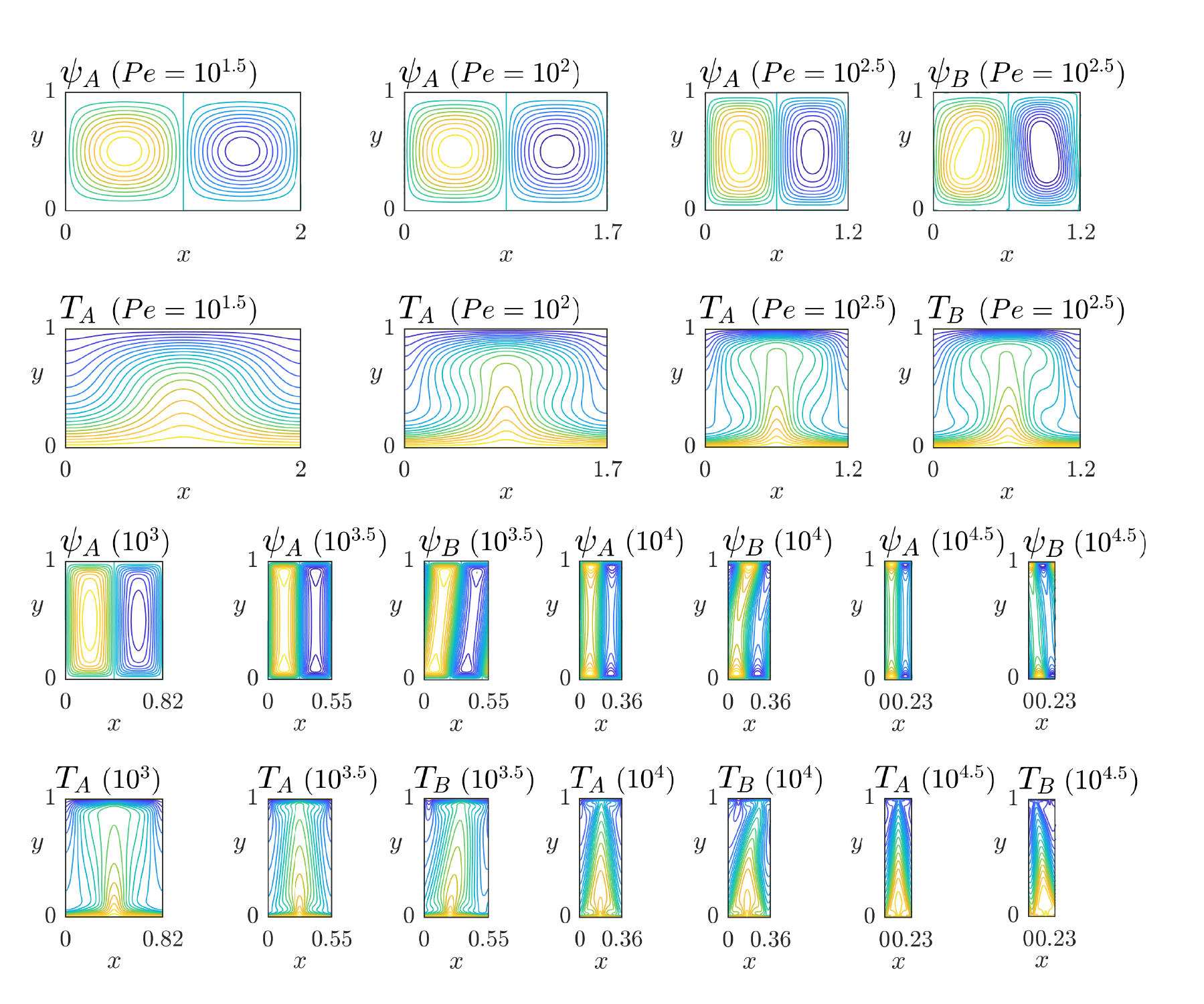}
    \caption{\footnotesize The progression of optimal flows and temperature fields as Pe increases from $10^{1.5}$ to $10^{4.5}$. Shown are the best flow and temperature fields among the optima, ($\psi_A, T_A$), and at four Pe an alternative optimum, ($\psi_B, T_B$), with the temperature field below the corresponding flow.}
    \label{fig:NewLowPowerOptima}
\end{figure}

In figure \ref{fig:NewLowPowerOptima} we show the progression of optimal flows and corresponding temperature fields as Pe increases from 10$^{1.5}$. Here the temperature field is the whole temperature field, not just the deviation from the conductive part as in figure \ref{fig:SmallPeOptima}. At Pe = $10^{1.5}$ the convective rolls are very close to those for Pe $\ll 1$ in 
figure \ref{fig:SmallPeOptima}A, and the local increase in the temperature where the flow is upward from the hot lower boundary at the $x$-midpoint can be seen. 

As Pe increases to $10^{4.5}$, the convective rolls become narrower horizontally, increasing the surface area where hot upwelling regions meet cold downwelling regions. The advantage for heat exchange is sufficient to outweigh the increase in viscous power dissipation due to the increase in $|\partial_x v|$ with convective roll narrowing. The viscous power dissipation is largest along the boundary, where $|\partial_y u|$ is particularly large. The best optimal flows are labeled $\psi_A$, and are the same as those in \cite{souza2020wall}. At Pe $= 10^{3.5}$ and above, small closed streamlines emerge close to the boundaries, showing a region of concentrated flow there. Despite this additional complexity we refer to the flows labeled $\psi_A$ collectively as ``simple roll" optima, in contrast to the alternative flow optima we show next. In four cases optima that are slight variations of the simple roll optima are shown, labeled $\psi_B$. At Pe $= 10^{2.5}$ the two convective rolls for $\psi_B$ are noticeably asymmetrical, while at Pe $= 10^{3.5}$ and above the two rolls are skewed in oblique directions. In these latter cases, we will show that there is little change in Nu ($\lesssim $1\%), so the skewed convective rolls are almost as efficient as the straight, simple rolls.

\begin{figure}
    \centering
    \includegraphics[width=7in]{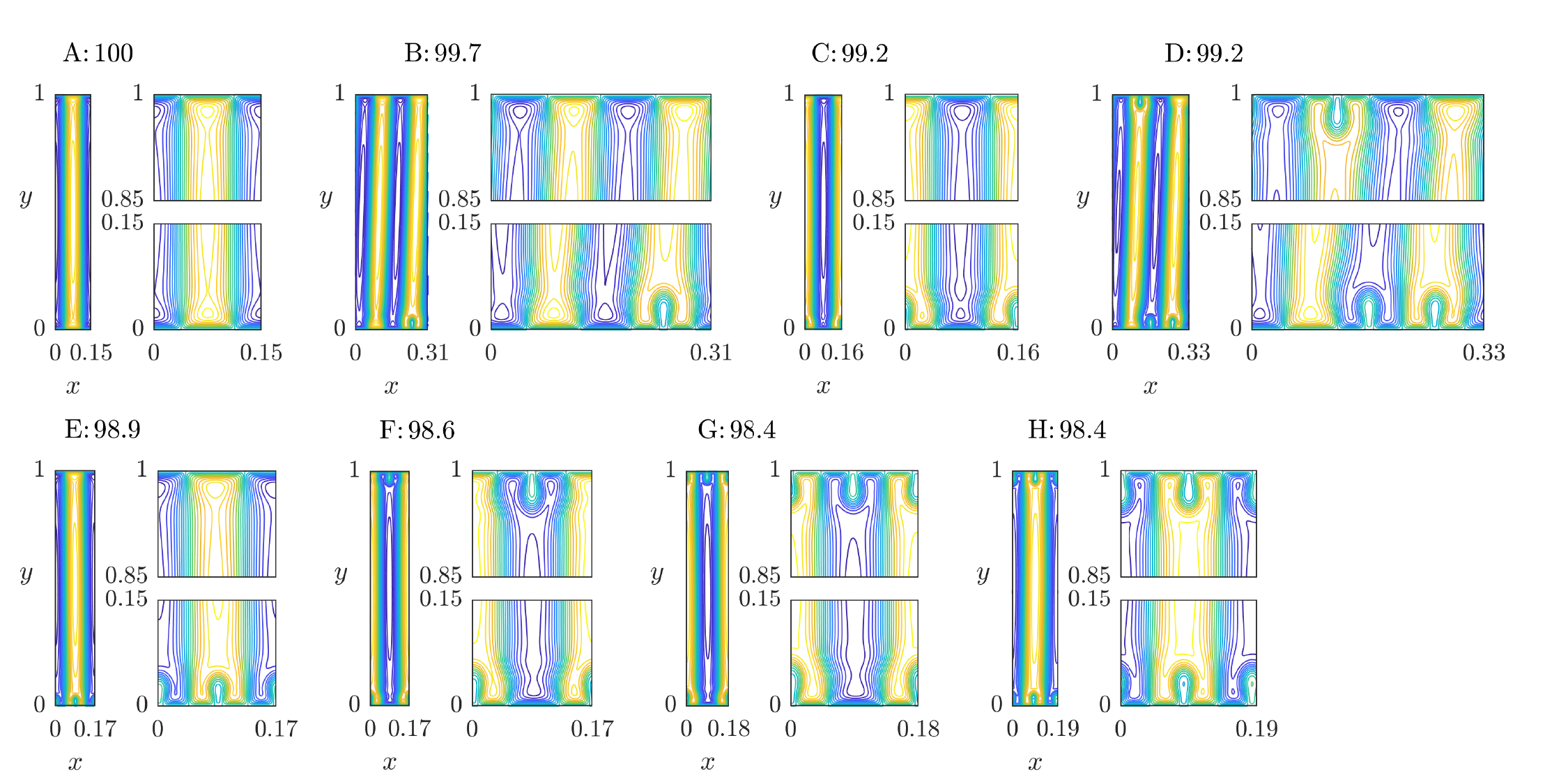}
    \caption{\footnotesize Optimal flow fields at Pe $= 10^5$.}
    \label{fig:NewPower10Optima}
\end{figure}

Between Pe $= 10^{4.5}$ and $10^{5}$, the class of optima changes dramatically with the emergence of branched flows. Examples of the best local optima at Pe $= 10^{5}$ are shown in figure \ref{fig:NewPower10Optima}. Each flow is shown by three panels, a tall narrow panel showing the streamlines over the whole domain, and to its right, two wider panels showing the streamlines close to the boundaries (within distance 0.15).
The flows are labeled at the top with their Nu value.
Flow A, with Nu = 100, is the simple convective roll optimum and is the global optimum. Flow B has two pairs of convective rolls, and one of the eight roll ends shows a branched structure near the bottom boundary. The flow is noticeably skewed, but by many comparisons of other examples of skewed and nonskewed optima, the skewness seems to have little effect on Nu. In flows C-H, the proportion of branched ends generally increases while Nu decreases, with some ties between neighboring flows. There are additional versions of these flows (not shown) with variations in the degree of skewness.

\begin{figure}
    \centering
    \includegraphics[width=7in]{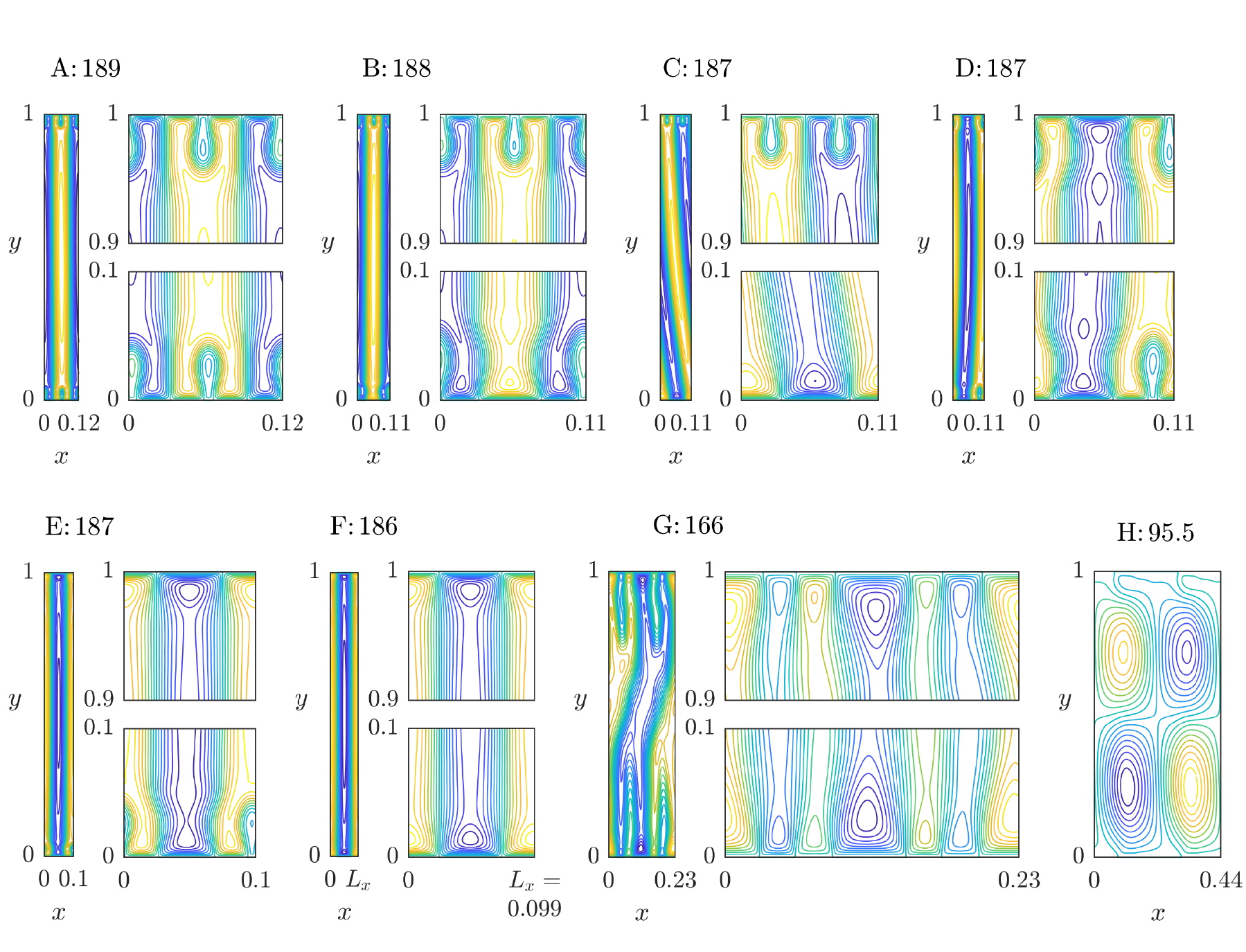}
    \caption{\footnotesize Optimal flow fields at Pe $= 10^{5.5}$.}
    \label{fig:NewPower11Optima}
\end{figure}

Optimal flows for Pe $= 10^{5.5}$, shown in figure \ref{fig:NewPower11Optima}, are similar, with slightly elongated branched regions compared to Pe $= 10^{5}$. Now the globally optimal flow (A) has branched ends. Moving from A to F, Nu decreases while the number of branched ends also decreases, the reverse trend of figure \ref{fig:NewPower10Optima}. The simple roll optimum is F with Nu = 186, within 2\% of the global optimum, 189 (A).
Optima G and H have much lower Nu, but are shown to illustrate the wide range of optima that may occur, including cases like G with branching but with Nu well below the global optimum. No inset is shown for H because the flow lacks fine structures near the walls and is surprisingly similar to the second best optimum for Pe $\ll 1$ (figure \ref{fig:SmallPeOptima}B). We hypothesize that G and H are just a small sample of a much larger space of optima with much lower Nu. The algorithm tends to find Nu close to the maximum in most cases, however, because at each iterative step, a backtracking line-search in the direction of the gradient is used, and it accepts a step when it gives a flow with higher Nu. As a result, the algorithm probably tends to wander away from local optima with relatively small Nu, such as G and H.

\begin{figure}
    \centering
    \includegraphics[width=6in]{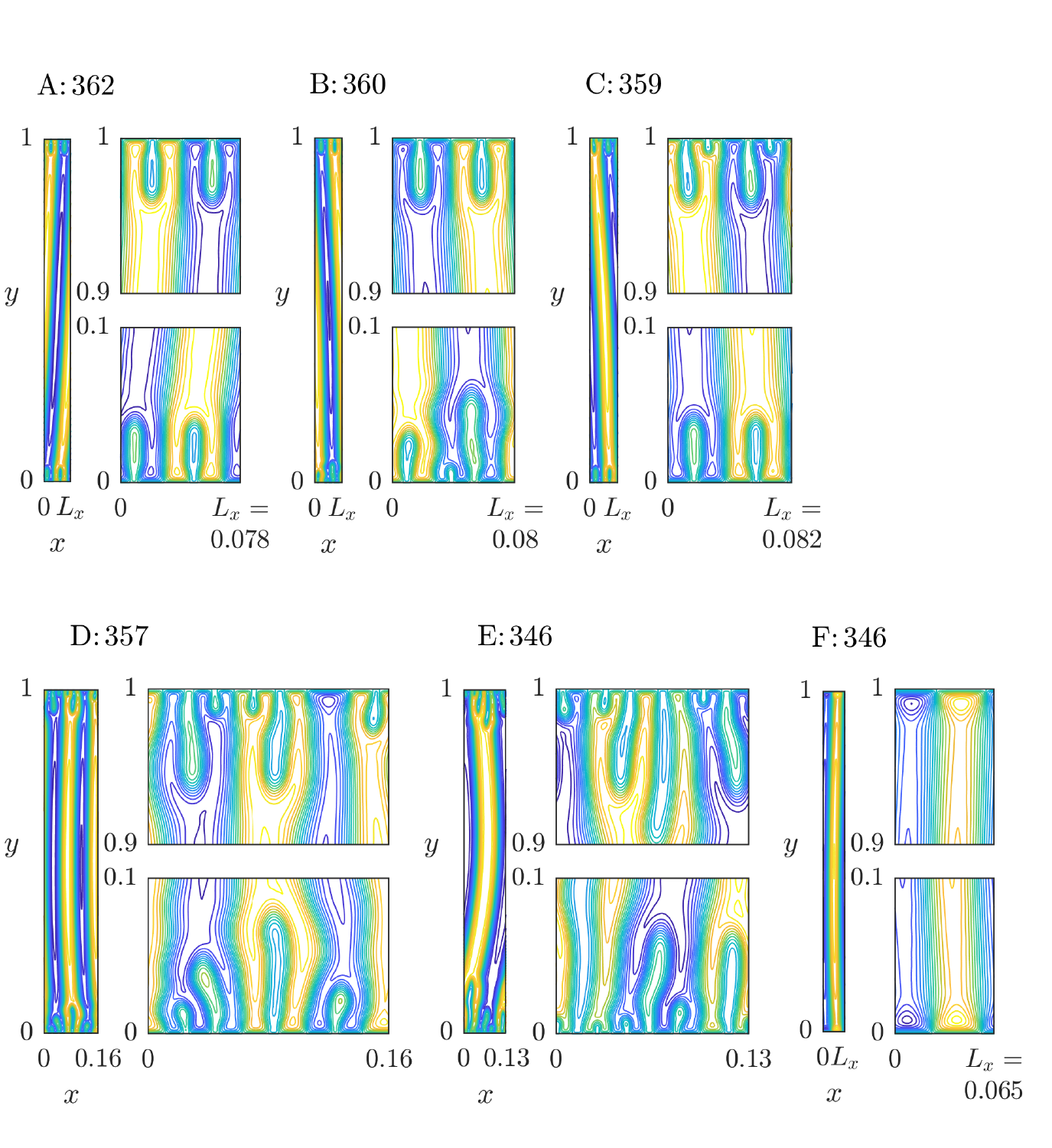}
    \caption{\footnotesize Optimal flow fields at Pe $= 10^{6}$.}
    \label{fig:NewPower12Optima}
\end{figure}

Figure \ref{fig:NewPower12Optima} shows optimal flows at Pe $= 10^6$. The branched tips are more elongated now, and small recirculation regions with closed streamlines are seen at their tips, similar to the simple roll optimum (flow F, with Nu = 346). There is also more skewness in the flows here, a general trend as Pe increases. Flows B-D have branched tips that undergo a second level of branching, a step towards the hierarchical branching flows of \cite{tobasco2017optimal}. Flow E is a case with three levels of branching near the top boundary. Although these flows are local optima, Nu decreases with the complexity of the branched structures, though there is also an increased gap between Nu for the global optimum (A), which has branching, relative to the simple roll optimum (F), with no branching.

\begin{figure}
    \centering
    \includegraphics[width=6in]{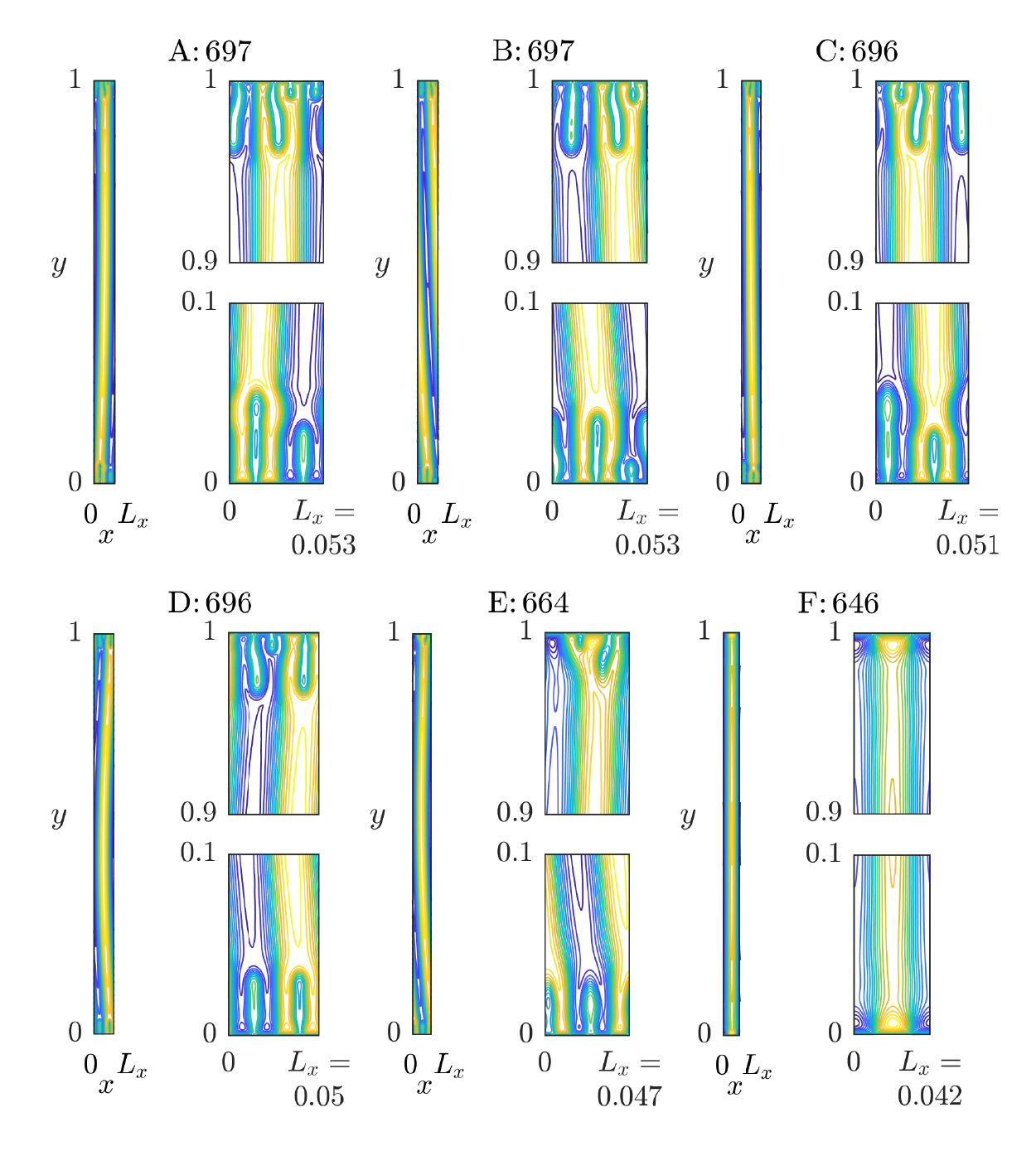}
    \caption{\footnotesize Optimal flow fields at Pe $= 10^{6.5}$.}
    \label{fig:NewPower13Optima}
\end{figure}

With Pe increased further, to $10^{6.5}$, figure \ref{fig:NewPower13Optima} shows a continuation of many of the same trends. Now the global optimum has two layers of branching. In all cases there is further elongation of the branched tips, which also adopt more complex shapes, with changes in width and direction near the walls (e.g. near the top walls in C and E). There is a steady decrease in the optimal $L_x$ with increasing Pe, which was identified with a power law behavior in \cite{souza2020wall} for the simple roll optimum.

\begin{figure}
    \centering
    \includegraphics[width=7in]{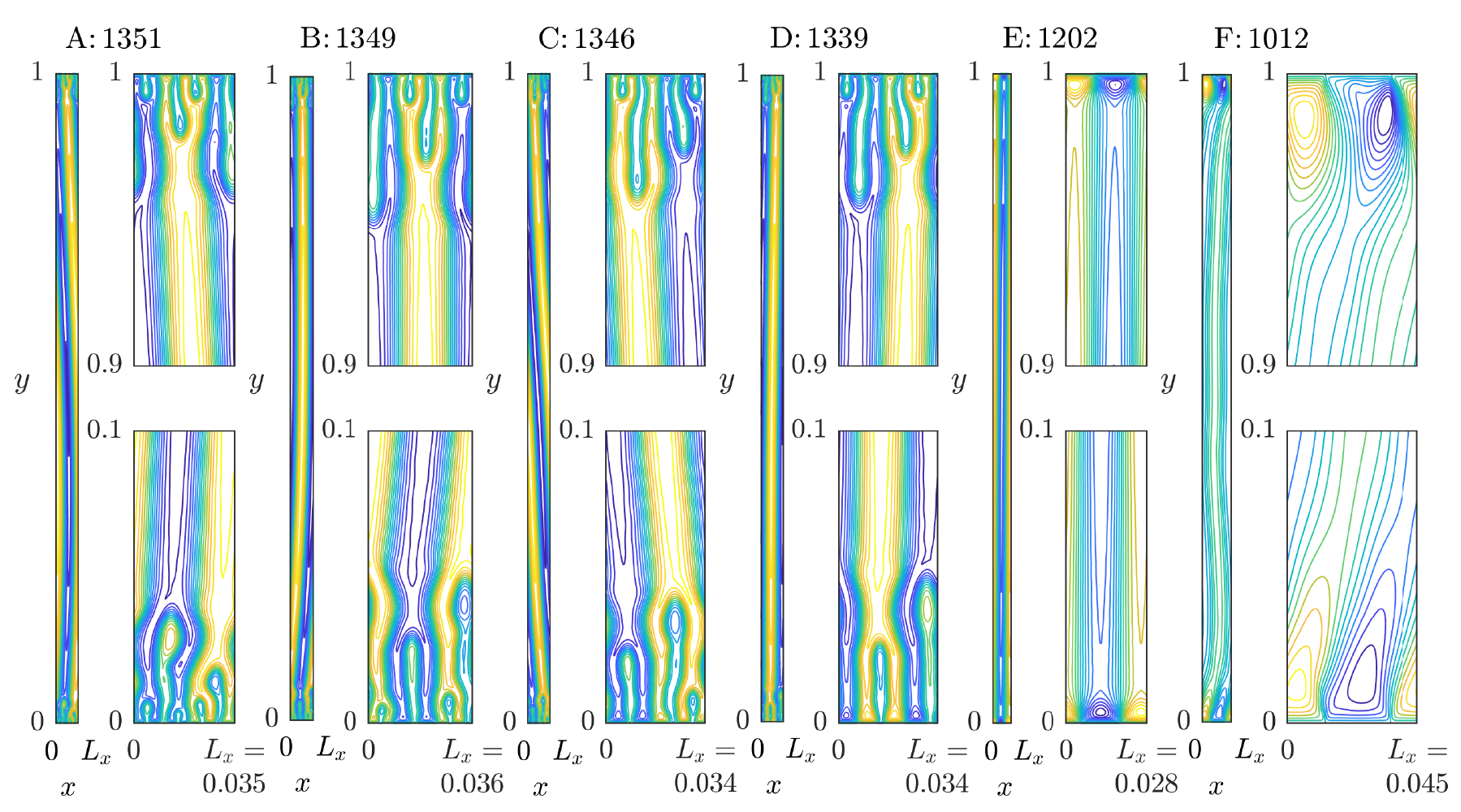}
    \caption{\footnotesize Optimal flow fields at Pe $= 10^{7}$.}
    \label{fig:NewPower14Optima}
\end{figure}

At Pe = $10^7$ (figure \ref{fig:NewPower14Optima}), the optimal flows (A-D) still have two layers of branching in some cases, but with increased elongation and waviness in the boundaries of the branches. The increase in Nu in flows A-D over that of the simple roll optimum (E) now exceeds 10\%. Flow F is an optimum with much lower Nu that resembles a skewed version of the simple convective roll. It also has larger and stronger eddies close to the walls, with less extreme values of the stream function in the interior, and therefore a weaker wall-to-wall flow.

\begin{figure}
    \centering
    \includegraphics[width=7in]{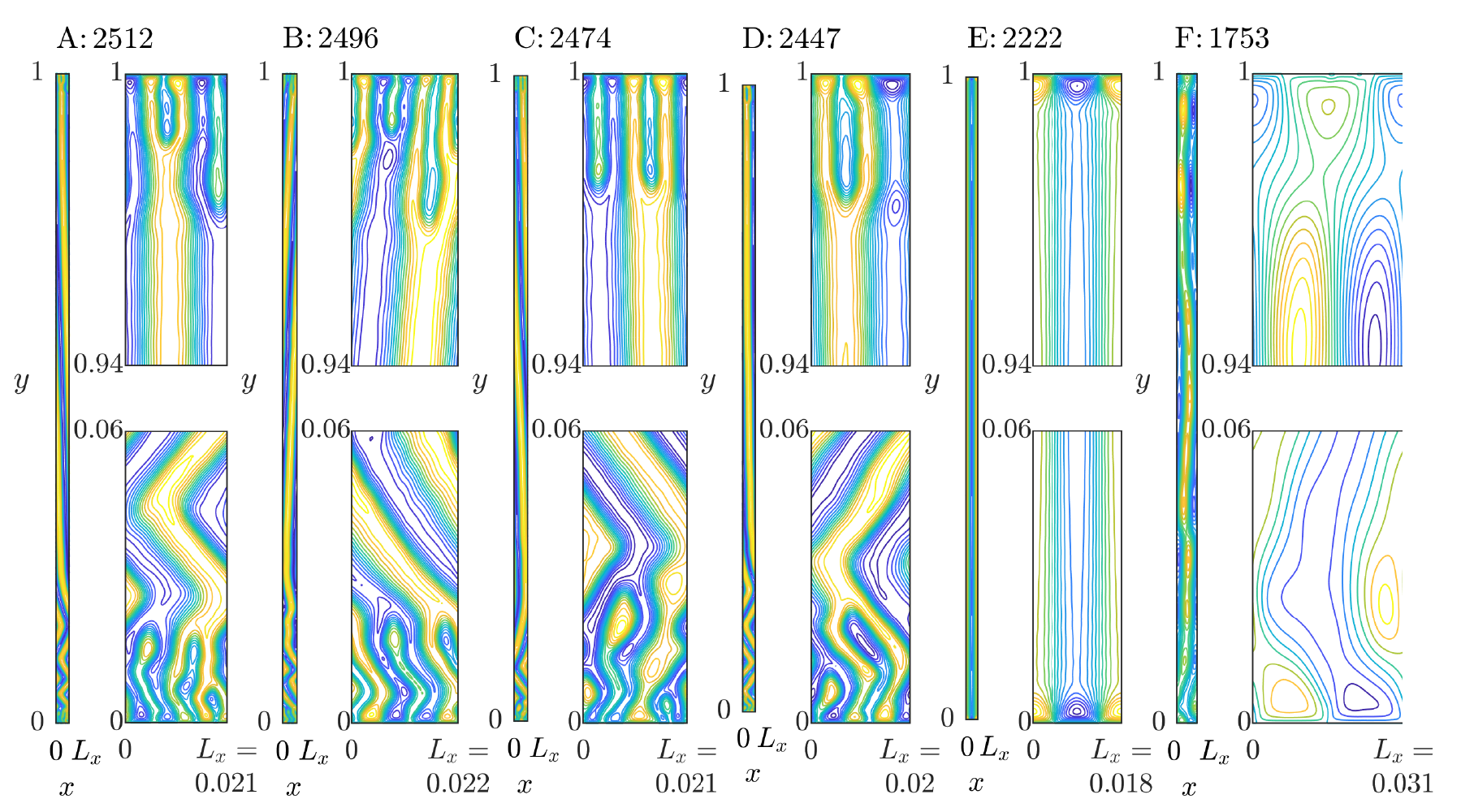}
    \caption{\footnotesize Optimal flow fields at Pe $= 10^{7.5}$.}
    \label{fig:NewPower15Optima}
\end{figure}

At the largest Nu studied here, Pe $= 10^{7.5}$, the flows again have one or two levels of branching and further elongation of the branches (figure \ref{fig:NewPower15Optima}). It is possible that the simultaneous optimization of $L_x$ and the flow inhibits the appearance of higher levels of branching, which would require larger $L_x$. Near the bottom boundary, flow C has three levels of very wavy branches shown by the blue streamlines (which can be seen by following the streamlines across the periodic boundary). There is generally a peculiar waviness of the flows near the bottom boundary. There is no asymmetry between the top and bottom boundaries in the model and no obvious asymmetry in the computations, except in the ordering of the unknowns for the matrix equations, i.e. (\ref{AdvDiffp}) for $\mathbf{T}$ and (\ref{q}) for $\mathbf{q}$. The unknowns are ordered starting at the lower left corner and proceeding rightward across a horizontal row and then upward row by row. We have not investigated the effect of the ordering of unknowns, but given that the discrete advection-diffusion operator becomes increasingly ill-conditioned as Pe increases, the solutions will eventually become noticeably inaccurate, and the waviness could be an early indication of this effect. Flow F is an example of an atypical optimum, with strong eddies distributed away from the walls and throughout the domain, and no branching near the walls. 

\begin{figure}
    \centering
    \includegraphics{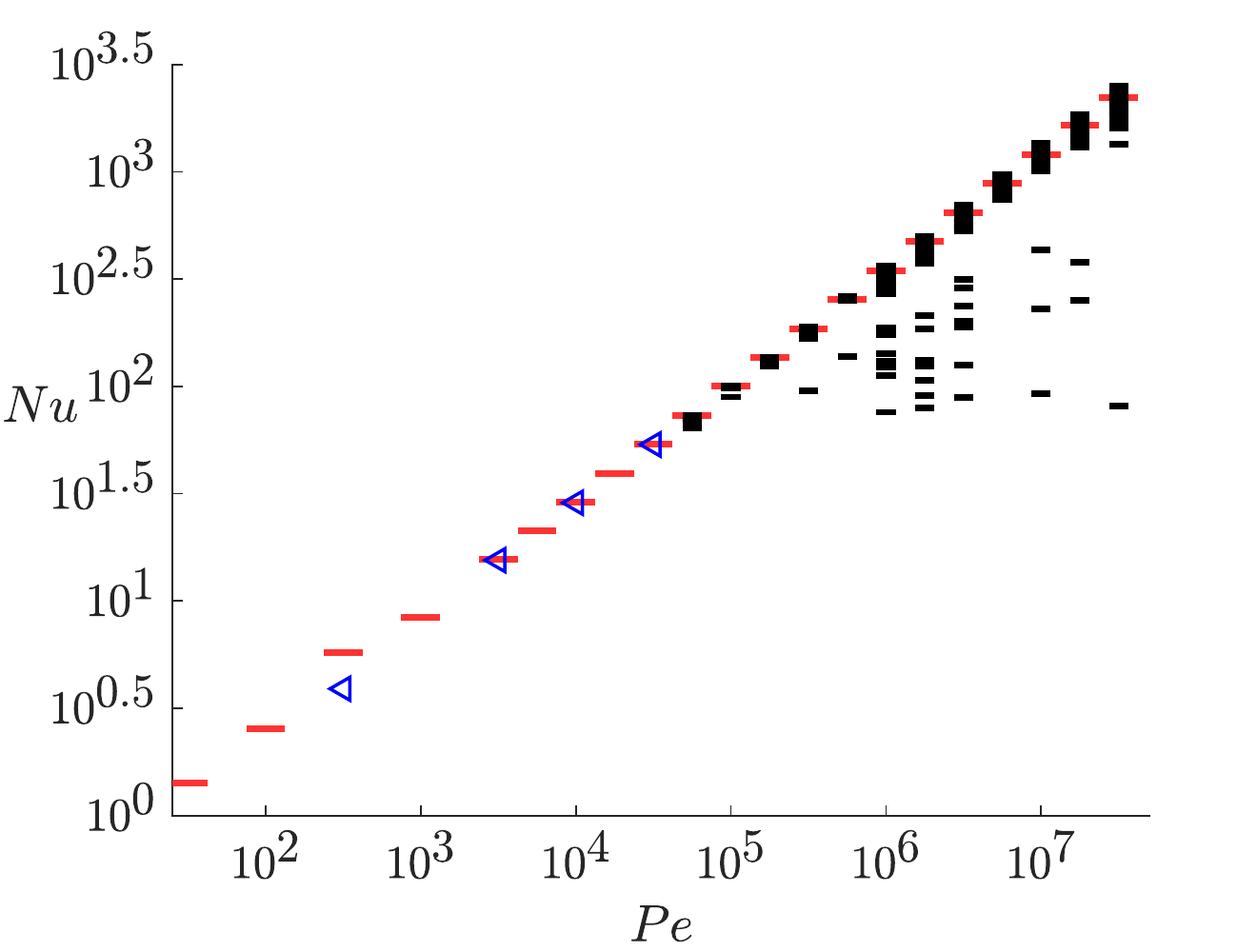}
    \caption{\footnotesize Nu versus Pe for optimal flow fields shown in this work and additional computed cases. The red hash marks give Nu for the simple roll optima. The black hash marks give Nu for other optima at Pe $\geq 10^{4.75}$. The blue triangles show Nu for $(\psi_B, T_B)$ in figure \ref{fig:NewLowPowerOptima}.}
    \label{fig:NewNuVsPe}
\end{figure}

In figure \ref{fig:NewNuVsPe}
we plot Nu for the flows shown thus far, and additional computed flows that were not shown. The simple roll optima Nu are shown with red hash marks centered at the corresponding (Pe, Nu) pair. The values for the four alternative flows in figure \ref{fig:NewLowPowerOptima} are shown by the blue triangles. The values for the alternative flows above the branching transition (for Pe $\geq 10^{4.75}$) are shown by the black hash marks, which can exceed Nu for the red hash marks for Pe $\geq 10^{5.25}$.

The simple convective roll solutions (red hash marks) fit Nu $\sim$ Pe$^{0.54}$, the same as in \cite{souza2020wall}. The upper envelope of the black marks, which correspond to certain branching flow optima, has a maximum slope of 0.575, noticeably larger than 0.54 but well below the 2/3 value seen in 3D \cite{motoki2018maximal} (and theoretically \cite{kumar2022three}) and found up to a logarithmic correction for self-similar 2D branching flows \cite{tobasco2017optimal}. At the upper end of the Pe range considered here, $10^7$--$10^{7.5}$, the slope drops from 0.575 to slightly below 0.54, indicating perhaps that the computations are less successful in approximating globally optimal flow(s) in this region.

\begin{figure}
    \centering
    \includegraphics[width=6.5in]{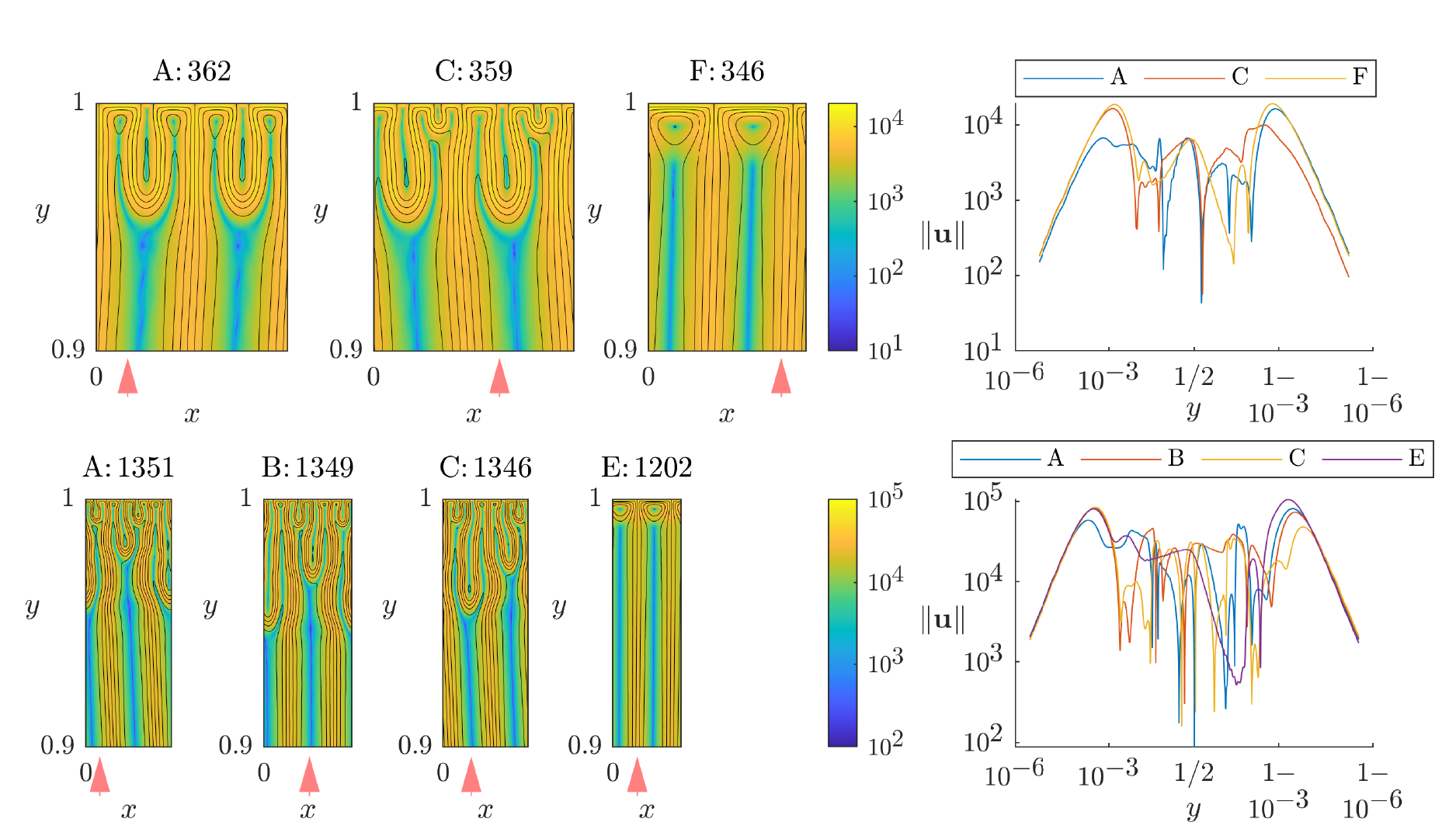}
    \caption{\footnotesize Color field plots of flow speed overlaid with streamlines (in black) for three optima from figure \ref{fig:NewPower12Optima} (in the top row) and
four optima from figure \ref{fig:NewPower14Optima} (in the bottom row), with labels at the top of each panel identifying the specific flow. At the right, the flow speed versus the vertical coordinate is plotted along one cross section for each panel, shown by the red arrows. }
    \label{fig:NewSpeedFig}
\end{figure}

So far we have presented the streamlines of the optimal flow configurations. The flow speed is proportional to the spacing between the streamlines, but since we limit the streamlines to 21 equally spaced values of the stream function for visibility, this gives only a rough estimate. Also, near the walls the streamlines are too close together to perceive their spacing. For a more accurate view, in figure \ref{fig:NewSpeedFig} we plot the flow speed as a color field near the top wall for three optima from figure \ref{fig:NewPower12Optima} (in the top row) and
four optima from figure \ref{fig:NewPower14Optima} (in the bottom row), with labels at the top of each panel identifying the specific flow. In each row, the color bar gives the speeds for all the color fields in that row. At the right, the flow speed versus the vertical coordinate is plotted along one cross section for each panel, shown by the red arrows. The cross-section is located at $x$ where the highest speed occurs in the whole flow domain, and the cross-section speed is plotted along the entire range $0 \leq y \leq 1$ even though the color fields are limited to $0.9 \leq y \leq 1$. On the horizontal axis, two separate logarithmic scales are used
for $0 < y \leq 1/2$ and $1/2 \leq y < 1$. The data are logarithmically spaced with respect to distance from $y = 0$ on the left half of the plot, and from $y = 1$ on the right half, to show the boundary layers near both walls.

For all seven flows considered here, the highest flow speed occurs at the boundary layer next to one of the walls. 
We define the boundary layer width $\delta$ for each flow as the distance from the flow speed maximum to the nearest wall.
For the top row, Pe $= 10^6$, the speeds reach a peak $\approx 2 \times 10^{4}$ over a boundary layer of width $\approx 10^{-3}$, while in the bottom row, Pe $= 10^7$, the speeds reach a peak $\approx 10^{5}$ over a boundary layer of width $\approx 4\times 10^{-4}$. The peak speeds and boundary layer widths are about 20\% higher for the simple convective roll than for the best optimum in the top row (A:362 and F:346) and about 40-60\% higher in the bottom row (A:1351 and E:1202), where Pe is larger.

For the simple convective rolls, we find that the maximum speed $\|\mathbf{u}\|_{max} \sim \mbox{Pe}^{0.75}$ and $\delta \sim \mbox{Pe}^{-0.49}$ for $10^5 \leq \mbox{Pe} \leq 10^{7.5}$.  The maximum power density is approximately the wall vorticity squared, $\approx \|\mathbf{u}_{max}\|^2/\delta^2 \sim \mbox{Pe}^{2.48}$, and grows faster than the average power density = Pe$^2$. For the optimal flows (panel A in figures \ref{fig:NewPower10Optima}-\ref{fig:NewPower15Optima}), we find $\|\mathbf{u}\|_{max} \sim \mbox{Pe}^{0.71}$ and 
$\delta \sim \mbox{Pe}^{-0.54}$. Both $\|\mathbf{u}\|_{max}$
and $\delta$ are smaller than for the simple convective roll, but the maximum power density is almost the same and has almost the scaling with Pe ($\sim \mbox{Pe}^{2.5}$).

\section{Conclusions and discussion \label{Conclusions}}

We have computed optimal steady flows for heat transfer in a 2D region between parallel walls, the ``wall-to-wall" problem \cite{hassanzadeh2014wall,motoki2018maximal,souza2020wall}. At Pe $\ll 1$, the optimal flows are squarish convection rolls, similar to the free-slip case \cite{hassanzadeh2014wall}. In contrast to previous studies, at large Pe we used an unconstrained optimization method with the adjoint method to compute gradients and the BFGS algorithm to maximize the Nusselt number.  Starting from 30-50 random initial flows, and increasing Pe from $10^{1.5}$ to $10^{7.5}$, we find a sequence of transitions. For moderate Pe, ($\in [10^{1.5}, 10^{4.5}]$), the algorithm converges to rectangular convection rolls (the ``simple convection roll"), or perturbations with skewed convection rolls that have nearly the same Nu as the simple convection roll in most cases. 

Between Pe = $10^{4.5}$ and $10^{4.75}$, optima with branched ends that have a simple U shape become prominent. They have Nu that is less than, but within 2\% of, that of the simple convective roll at Pe = $10^{5}$. Here Nu decreases with the degree of branching. At Pe = 
$10^{5.5}$, Nu instead increases with the degree of branching, and the optimal solution has bilaterally symmetric rolls with U-shaped branched ends. The simple convective roll, without branching, has Nu within 2\% of it. Very different optima with much lower Nu are also found, indicating a much larger space of local optima. At Pe = $10^6$, optima with two and three layers of branching are found, though the best optimum again has a single layer of U-shaped branching at the walls. As Pe increases, there is a general trend towards more vertically elongated branching as well as smaller $L_x$, perhaps for the same reason--to increase lateral conduction across convective rolls. There is also the presence of small closed eddies at the branched ends, as occurs for the simple convective roll. The branching patterns become more complex and deviate more from symmetry across vertical planes. These trends continue at Pe = $10^{6.5}$, $10^{7}$, and $10^{7.5}$. 

The scaling exponent for Nu with respect to Pe is as large as 0.575, larger than 0.54 for the simple convective roll, but well short of the 2/3 value seen in the 3D computations of \cite{motoki2018maximal} and in the analytical solutions of \cite{tobasco2017optimal}, up to logarithmic corrections. We also studied the scaling of the flow speed and boundary layer thickness near the wall, defined as the distance from the wall to the global maximum speed, which always occurs very close to the wall, at the end of a monotonic rise from zero velocity at the wall.
For the branched flows, the maximum speed and the boundary layer thickness are both somewhat smaller than for the simple convective roll at the same Pe, and so are the power law exponents. However, the maximum power density, the squared ratio of the maximum speed and boundary layer thickness, is very close in both cases, with scalings close to $Pe^{2.5}$, compared to the $Pe^{2}$ spatial average.

Two-dimensionality makes the problem more tractable computationally than the 3D case \cite{motoki2018maximal}, though this is balanced somewhat by the need to compute at higher Pe to study the branching transition and the asymptotic scaling of Nu with Pe, still unclear at the largest Pe used here, $10^{7.5}$. In \cite{motoki2018maximal} a Nu $\sim \mbox{Pe}^{2/3}$ scaling is already clear at Pe = $10^4$. At higher Pe the optimal flows and corresponding temperature fields have sharper gradients, which may inhibit the convergence to optimal solutions.

\appendix

\section{Other expressions for the Nusselt number \label{OtherNu}}

Writing (\ref{AdvDiff}) in divergence form and integrating over a horizontal period,
\begin{equation}
0 = \int_0^{L_x} \nabla \cdot (\mathbf{u} T - \nabla T)\, dx =  \int_0^{L_x} \partial_x (uT - \partial_xT) + \partial_y (vT - \partial_yT)\, dx =  \partial_y \int_0^{L_x}  (vT - \partial_yT) \,dx
\end{equation}
using the periodicity of the flow and temperature for the last equality. Thus 
\begin{equation}
\frac{1}{L_x}\int_0^{L_x}  (vT - \partial_yT)\, dx \label{flux}
\end{equation}
is constant in $y$. Evaluating (\ref{flux}) at $y = 0$ or $y = 1$ (where $v$ = 0), we have the rate of heat transfer from the hot boundary or the rate of heat transfer to the cold boundary, both equal to Nu. Integrating (\ref{flux}) from $y$ = 0 to 1, we have
\begin{equation}
\mbox{Nu}= 1 + \frac{1}{L_x}\int_0^1\int_0^{L_x}  vT \,dx \,dy.
\end{equation}
\cite{tobasco2017optimal} shows that Nu has a minimum value of 1 for the purely conductive temperature field ($T = 1-y$) that occurs with $\mathbf{u} = \mathbf{0}$, and is strictly greater than 1 for any other $\mathbf{u}$ (and the corresponding $T$).

\section{Numerical details \label{Numerical}}

The subspace of $\langle T_0(2y-1), \ldots, T_{k+4}(2y-1) \rangle$ that obey the no-slip conditions at $y = 0$ and 1 consists of $\sum A_k T_k(2y-1)$ such that
\begin{align}
    \sum A_k T_k(-1) = \sum (-1)^k A_k = \mathbf{v}_1^T\mathbf{A} = 0 \\
    \sum A_k T_k(1) = \sum A_k = \mathbf{v}_2^T\mathbf{A} = 0 \\
    \sum A_k T_k'(-1) = \sum (-1)^k k^2 A_k = \mathbf{v}_3^T\mathbf{A} = 0 \\
    \sum A_k T_k'(1) = \sum k^2 A_k = \mathbf{v}_4^T\mathbf{A} = 0 \\
\end{align}
In the basis $\{ T_0(2y-1), \ldots, T_{k+4}(2y-1)\}$, $T_{k+4}$ is the unit basis vector $\hat{\mathbf{e}}_{k+4}$. We subtract its orthogonal projection onto
the four-dimensional subspace \{$\mathbf{v}_1, \ldots, \mathbf{v}_4$\} to obtain
the coefficients of $Y_k(y)$ in the basis $\{ T_0(2y-1), \ldots, T_{k+4}(2y-1)\}$.
This determines $Y_k(y)$.

\begin{acknowledgments}
This research was supported by the NSF-DMS Applied Mathematics program under
award number DMS-2204900.
\end{acknowledgments}


\end{document}